**Article title**
Are they just delegating?
Cross-Sample Predictions on University Students' & Teachers' Use of AI.

**Author names**
Fabian Albers [a], Sebastian Strauß [b], Nikol Rummel [b, c], Nils Köbis [a]

**Affiliations**
[a] Research Center for Trustworthy Data Science and Security, University of Duisburg-Essen
Bismarckstraße 120, 47057 Duisburg, Germany.
E-Mail: fabian.albers@uni-due.de, nils.koebis@uni-due.de
[b] Ruhr-University Bochum
Universitätsstraße 150, 44801 Bochum, Germany.
E-Mail: sebastian.strauss@rub.de, nikol.rummel@rub.de
[c] Center for Advanced Internet Studies
Konrad-Zuse-Straße 2a, 44801 Bochum, Germany.
E-Mail: nikol.rummel@cais-research.de

**Corresponding author**
Fabian Albers
Research Center for Trustworthy Data Science and Security, University of Duisburg-Essen
Bismarckstraße 120, 47057 Duisburg, Germany.
E-Mail: fabian.albers@uni-due.de

**Declaration of competing interests**
The authors declare that they have no known competing financial interests or personal relationships that could have appeared to influence the work reported in this paper.

*This research did not receive any specific grant from funding agencies in the public, commercial, or not-for-profit sectors.*

**Abstract**

Mutual trust between teachers and students is a prerequisite for effective teaching, learning, and assessment in higher education. Accurate predictions about the other group's use of generative artificial intelligence (AI) are fundamental for such trust. However, the disruptive rise of AI has transformed academic work practices, raising important questions about how teachers and students use these tools and how well they can estimate each other's usage. While the frequency of use is well studied, little is known about *how* AI is used, and comparisons with similar are rare. This study surveyed German university teachers ($N$ = 113) and students ($N$ = 123) on the frequency of AI use and the degree of delegation across six identical academic tasks. Participants also provided incentivized cross-sample predictions of the other group's AI use to assess the accuracy of their predictions. We find that students reported higher use of AI and greater delegation than teachers. Both groups significantly overestimated the other group's use, with teachers predicting very frequent use and high delegation by students, and students assuming teachers use AI similarly to themselves. These findings reveal a perception gap between teachers' and students' expectations and actual AI use. Such gaps may hinder trust and effective collaboration, underscoring the need for open dialogue about AI practices in academia and for policies that support the equitable and transparent integration of AI tools in higher education.

# 1. Introduction

Trusting relationships between university teachers and students are a prerequisite for learning, teaching, and assessment in higher education (Carless, 2009; Luo, 2025; Mayhew et al., 2016). However, since trust is largely based on accurate predictions of each other's behavior (Dietz, 2011; Six & Latusek, 2023), teacher-student relationships may need to be recalibrated due to the disruptive emergence of generative AI in higher education (Kramm & McKenna, 2023; Luo, 2025).

Consider the example of grading from the teacher's perspective. If teachers underestimate students' use of AI, they may overtrust students and choose not to require students to explicitly indicate use of AI tools. This can lead to problems with performance assessment due to the uncertainty of authorship (Carless, 2009). If teachers overestimate students' use of AI, they may undertrust students and introduce very strict rules for the use of AI tools. This can create an atmosphere of surveillance and distrust, discouraging students from exploring AI's learning-enhancing features (Luo, 2025; Ross & Macleod, 2018).

Now consider an example from the students' perspective. If students underestimate teachers' use of AI, they may overtrust learning materials and overlook possible AI-generated errors, such as hallucinations. If students overestimate teachers' use of AI, they may undertrust the course materials, become overly critical, feel less motivated, or feel more justified in using AI themselves to complete assignments. This point is echoed in a recent report in The New York Times, outlining how American students criticized the hypocrisy of professors using AI in teaching while they are not allowed to do so (Hill, 2025). In both cases, misjudgments of the other's use of AI can shape teachers' and students' willingness to use AI for academic purposes and may undermine the mutual trust that is essential for effective teaching and learning.

Accurate predictions about each other's AI use are, therefore, a prerequisite for fruitful teacher-student relationships in higher education in the age of AI. Yet, there is a lack of empirical evidence on (a) the assessment of AI use among students and teachers using equivalent measurement instruments, and (b) their perceptions and predictions of each other's use. To address this gap, we conducted online surveys with teachers (*N* = 113) and students (*N* = 123) at German universities. The two surveys comprised questionnaires that evaluated both teachers' and students' own AI use, specifically, the frequency of use and degree of delegation across several key tasks. Additionally, we provide some of the first insights into cross-sample predictions. That is, what teachers believe about students' use of AI and vice versa. For that, we draw on state-of-the-art incentivized measures from behavioral economics to assess such beliefs (Voslinsky & Azar, 2021).

## 2. Background

### 2.1 AI Use in Higher Education

Generative AI is transforming work practices across many fields, with higher education being among the sectors where adoption has occurred most rapidly. This fast technological change has sparked ongoing discussion about the role of AI tools in academic contexts (Bittencourt et al., 2020; Díaz & Nussbaum, 2024). Similar to other contexts where people delegate tasks to AI (Köbis et al.), a wide range of academic tasks, whether in teaching or learning, can be supported or delegated to AI systems to varying degrees in universities (Cukurova, 2025). Understanding delegation to AI in the academic context is essential because AI in education offers numerous benefits, including efficiency and enhanced learning opportunities (Giannakos et al., 2025; Huang et al., 2023). Yet it also poses risks when it affects skill development, authorship (Imran & Almusharraf, 2023), or trust between teachers and students (Luo, 2025). Therefore, it is crucial to understand *if*, *for which tasks*, and *to what extent*, this delegation occurs in academic contexts.

Against this backdrop, research on the use of AI in higher education is rapidly expanding. Studies on university teachers' adoption of AI are often exploratory, focusing on their experiences, attitudes, perceived opportunities, and concerns (Lau & Guo, 2023; Mohammadi et al., 2026). In contrast, research on university students already provides more extensive insight into their AI adoption in academic contexts (Reiter et al., 2025; Stöhr et al., 2024; Von Garrel & Mayer, 2023).

On the teachers' side, results from a recent 20-country survey indicate that 41% of university teachers use AI at least monthly, while 32.7% report never having used it (Mohammadi et al., 2026). These results align with an exploratory study that interviewed 20 university teachers from nine countries about their current use of AI and their plans to integrate it into teaching (Lau & Guo, 2023). The findings suggest that almost half of the teachers had used AI, with stated use ranging from minimal engagement to full integration into their teaching practices (ibid.). Results from the European context present a similar picture; A recent study on the teaching practices of Bulgarian university teachers found that 42.53% of participants reported never having used AI (Kiryakova & Angelova, 2023). Findings from a similar study focusing on the relationship between teacher perceptions and AI use even suggest that the majority of German STEM teachers have never used AI in an academic context (Beege et al., 2024). For academic applications of AI, teachers reported using it for tasks such as information research, idea generation, providing feedback, and writing or visualizing teaching content and materials (Kiryakova & Angelova, 2023; Mohammadi et al., 2026). Overall, these findings suggest that while a significant proportion of teachers across various contexts have begun integrating AI into specific academic tasks, non-use remains prevalent, highlighting substantial variability in adoption levels and purposes.

On the students' side, data from multiple surveys conducted across various Western, Educated, Industrialized, Rich, and Democratic (WEIRD) (Henrich et al., 2010) countries indicate that the vast majority of university students have used AI in academic contexts, with many reporting frequent use (Faraon et al., 2025; Hornberger et al., 2025). In the German context, recent studies indicate that two-thirds of students have used AI in their studies (Reiter et al., 2025; Von Garrel & Mayer, 2023). A more detailed breakdown reveals that 25.2% of students use AI-based tools very frequently, while an additional 47.8% report using them only rarely or occasionally (Von Garrel & Mayer, 2023). Students report using AI most frequently for tasks such as information research, literature research, translation, idea development, text analysis, and data visualization, as well as processing and creation (Reiter et al., 2025; Von Garrel & Mayer, 2023). However, students' frequency of AI use, particularly across different academic tasks, varies notably by field of study (ibid.). Overall, these findings suggest that AI adoption among students is relatively common but varies in intensity and across different disciplines.

Based on these findings, we hypothesize:

H1: The self-reported frequency of AI use for academic tasks is higher for university students than for teachers.

While previous work offers rich insights into AI adoption in terms of the frequency of university teachers' and students' use of AI, there is a lack of research examining *how* these tools are applied in academic contexts. From the students' perspective, it is particularly important to consider the extent to which academic tasks are delegated to AI tools, as excessive reliance can bypass essential cognitive processes that are required to construct, connect, and apply knowledge (Chi & Wylie, 2014; Hou et al., 2025). It is particularly relevant to distinguish between cases where AI use fosters engagement, such as discussing ideas or refining phrasing, and cases where entire tasks are handed over to AI, potentially bypassing deeper learning processes.

Our hypothesis is based on the assumption that a higher frequency of AI use also entails a higher degree of delegation to AI. We therefore hypothesize that:

H2: The self-reported degree of AI delegation for academic tasks is higher for university students than for teachers.

**2.2 Cross-Sample Predictions in the Age of AI**

Accurate predictions of each other's behavior are central to beliefs about trustworthiness (Dietz, 2011). The same applies in higher education, where mutual trust between teachers and students can be defined as "an individual's or group's willingness to be vulnerable to another party based on the confidence that the latter party is benevolent, reliable, competent, honest and open" (Hoy & Tschannen-Moran, 1999, p. 189). Trusting relationships between university teachers and students are based on "confident positive expectations" (Dietz, 2011, p. 219) that largely stem from prior interactions. However, in contexts where prior experiences are scarce, as in the rapidly emerging use of AI in higher education, teachers and students may struggle to form accurate predictions about each other's behavior. These inaccurate predictions may lead to a lack of trust, also referred to as mistrust (Six & Latusek, 2023). In this way, the disruptive emergence of AI use in higher education may pose a risk to trusting relationships between teachers' and students' relationships (Luo, 2025; Macfarlane, 2021; Ross & Macleod, 2018), which are widely regarded as a prerequisite for effective learning, teaching, and assessment in higher education (Carless, 2009; Felten et al., 2023; Mayhew et al., 2016).

On the teachers' side, findings of a recent qualitative study show that nearly all the interviewed university teachers stated that they did not know the extent of students' AI use (Lau & Guo, 2023). This finding aligns with a growing body of literature that argues for a growing atmosphere of distrust between teachers and university students, reflected in the increased reliance on anti-plagiarism software and learning analytics to monitor engagement (Macfarlane, 2021). As argued, this technologically mediated plagiarism detection is situated within a broader culture of surveillance and distrust in contemporary higher education, impacting relationships among teachers, students, and institutions (Ross & Macleod, 2018).

On the students' side, a recent qualitative interview study found relatively low levels of trust in teachers when navigating assessment in the age of generative AI (Luo, 2025). Students reported fears of being wrongly accused of AI-assisted cheating or penalized for admitting to AI use (ibid.). The authors argue that this general suspicion creates barriers to open communication and reduces their willingness to take the risks necessary for trust-building (ibid.).

To examine whether prediction accuracy regarding the other group's use of AI in academic contexts contributes to this current mistrust, we address our third research question: *How* *accurately do university students predict teachers' AI use in terms of frequency and degree of delegation, and vice versa?* — by collecting incentivized predictions of the other group's frequency of use and degree of delegation across six identical academic tasks completed by both groups. We base our hypotheses on the idea that effective teaching involves continuously monitoring and responding to students' needs (Hattie, 2009). These practices provide teachers with more opportunities to observe and interpret the behavior of large and diverse student groups than students have to observe teachers, whose practices outside the classroom often remain inaccessible to them. We therefore hypothesize that:

H3.1: University teachers' predictions of students' self-reported frequency of AI use for academic tasks will be more accurate than students' predictions of teachers' self-reported frequency of AI use for academic tasks.

H3.2: University teachers' predictions of students' self-reported degree of AI delegation for academic tasks will be more accurate than students' predictions of teachers' self-reported degree of AI delegation for academic tasks.

## 3. Methods

We conducted online surveys with two participant groups: university teachers and university students. Each group completed a role-specific questionnaire assessing their own *frequency* of using AI for a range of typical academic tasks (e.g., information research, writing, visualizing). They also indicated the *degree* to which they delegate these tasks. Moreover, each group made incentivized cross-sample predictions about the other group's frequency and degree of delegation for the same tasks. *Figure 1* illustrates the study design.

**Figure 1.**

*Conceptual diagram of the present study.*

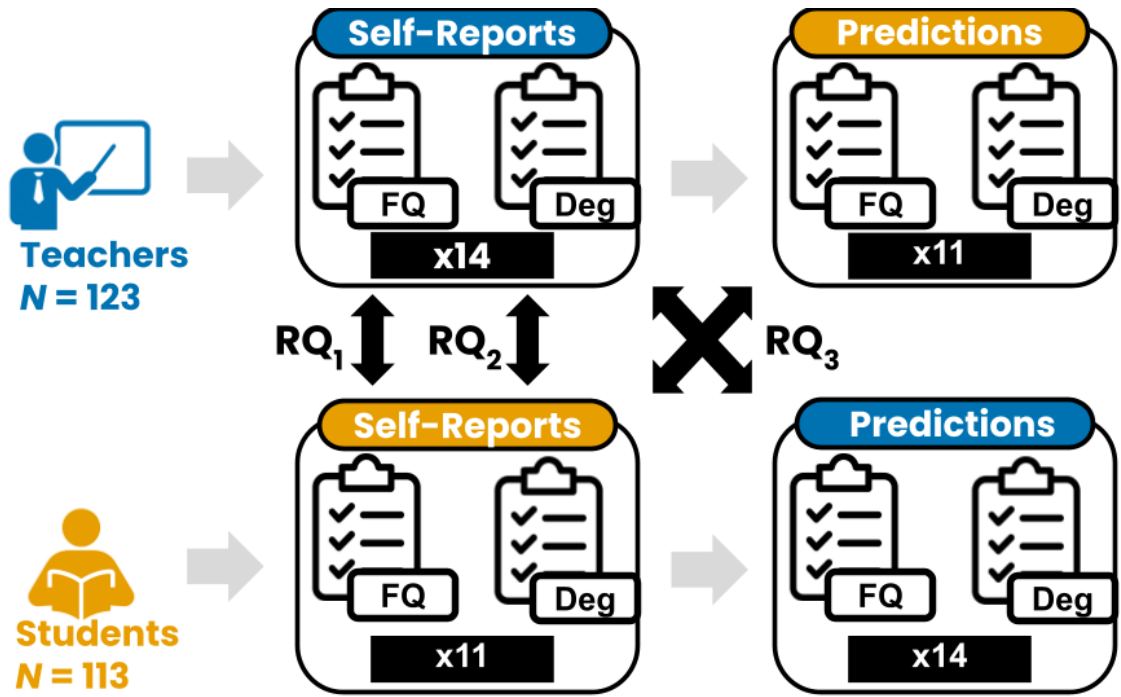

*Note*. frequency of AI use (FQ); degree of AI delegation (Deg)

### 3.1 Open Research Statement

This study was pre-registered prior to data collection at AsPredicted (Registration #226484). The pre-registration specifies all hypotheses, procedures, and planned analyses. Unless otherwise noted, the study followed this plan. Additional information, including documented deviations, full study materials, anonymized data, and analysis scripts, is available in the Supplementary Information (SI) hosted on the Open Science Framework. The study received a positive ethics vote from the Ethics Committee of "blinded for peer review" (Approval No. EPE-2025–004), confirming compliance with institutional and national ethical standards for research with human participants.

### 3.2 Participants

The samples consisted of teachers (*N* = 113) and students (*N* = 123) from German universities, representing a diverse range of disciplines and fields of study. An a priori power analysis indicated that 100 participants per condition were required to detect the hypothesized medium-sized effects (see AsPredicted #226484). According to the preregistered plan, teachers were recruited through professional networks, university mailing lists, and targeted email invitations, beginning on May 5, 2025, over an eight-week period. In total, *N* = 1091 teachers began the online study. After applying preregistered exclusion criteria of (a) not meeting eligibility requirements, meaning teaching/being enrolled as a student at a German university, (b) providing incomplete data, or (c) submitting duplicate responses, the final teacher sample comprised *N* = 123 ($M_{age}$= 39.2 ; $SD_{age}$= 10.9).

Because student recruitment through university channels did not reach the target sample size determined by the power analysis, a second sample was obtained via Prolific (see AsPredicted #240922). This second wave of data collection yielded *N* = 129 student participants. After applying the same preregistered exclusion criteria, the final student sample comprised *N* = 113 ($M_{age}$= 26.7; $SD_{age}$= 6.35). This is the sample used in all subsequent analyses. For transparency, we also report analyses conducted with the initial sample in the SI and indicate whenever the results differ substantially.

### 3.2.1 Remuneration

As a baseline pay, teachers could choose a charitable organization to which we donated 5€ in their name. Students recruited via Prolific received a baseline payment of €4. Furthermore, both groups could earn performance-based bonuses dependent on their prediction accuracy. For the prediction of the frequency of AI use, we added a bonus of €5 to teachers' donations if their average prediction across all items matched students' self-reported average. Vice versa, students received a bonus payment of €1.

For the prediction of the degree of AI delegation, we added a bonus of 5€ if teachers' average prediction across all items was within ±5% of students' self-reported average, and 9€ if it exactly matched the actual average (±0%). Vice versa, students received bonus payments of 1€ (±5%) and 9€ (±0%), respectively.

### 3.3 Procedure

After providing informed consent, participants completed a survey consisting of two main parts. In the first part, teachers and students provided their self-reported (a) frequency of AI use and (b) degree of AI delegation across 11 (students)/ 14 (teachers) tasks. In the second part, participants provided predictions about the other groups' (a) frequency of AI use and (b) degree of AI delegation across the corresponding academic tasks. Following this, participants had the option to complete a contact questionnaire to receive feedback about their performance later. Finally, all participants were thanked and debriefed.

### 3.4 Measures

The survey comprised (a) self-reports and (b) cross-sample predictions about the frequency and degree of delegation of AI for academic tasks. For each measure, we used a set of predefined academic tasks, 14 for teachers and 11 for students, derived from prior research (Beege et al., 2024; Reiter et al., 2025; Von Garrel & Mayer, 2023). For data analysis, we focus on the six overlapping academic tasks, which are relevant to both groups (information research, literature research, writing, coding, translating, and visualizing). For further exploratory analysis, see SI. We conducted the studies in German.

*Self-reported frequency of AI use*. We measured participants' self-reported frequency of AI use with a single five-point scale item, asking: "How often do you use AI for this task?" (1 = never, 2 = rarely, 3 = sometimes, 4 = often, 5 = always) for each academic task.

*Self-reported degree of AI delegation*. We measured the perceived degree of AI delegation with a single continuous scale item implemented as an on-screen slider, asking: "How big is your/ AI's contribution to the outcome of this task?" Participants rated their relative contribution of themselves versus AI to the task on a 100-point scale ranging from "100% me" to "100% AI".

*Predicted frequency of AI use*. We measured participants' predictions of the other group's frequency of AI use with a single five-point scale item, asking: "How often do you think [teachers/students] use AI for this task?" (1 = never, 2 = rarely, 3 = sometimes, 4 = often, 5 = always) for each academic task.

*Predicted degree of AI delegation*. We measured participants' predictions of the other group's degree of AI delegation with a single continuous scale item implemented as an on-screen slider, asking: "How big is the contribution of [teachers/students]/AI to the outcome of this task?" Participants rated the other group's relative contribution of teachers/students versus AI to the task on a 100-point scale ranging from "100%(teachers/students)" to "100% AI".

*Prediction accuracy.* We calculated prediction accuracy as the absolute value of the difference between each participant's prediction about the other group and that group's mean self-report for every academic task. We did this separately for (a) frequency of AI use and (b) degree of AI delegation. Taking the absolute value means we treat overestimates and underestimates the same. Lower values show higher accuracy.

## 4. Results

**RQ1: How frequently do university students/teachers use AI for academic tasks?**

We conducted a linear mixed-effects regression to examine differences in self-reported frequency of AI use between university students and teachers, as illustrated in Figure 2. The model included group (students vs. teachers), task type, and their interaction as fixed effects. A random intercept for participants accounted for individual differences in self-reported frequency of AI use, with $\sigma^2 = 0.29$ (*SD* = 0.54), and a residual variance of $\sigma^2 = 0.96$ (*SD* = 0.98), reflecting within-participant variability across academic tasks.

**Figure 2.**

*Self-reported Frequency of AI Use of University Teachers and Students.*

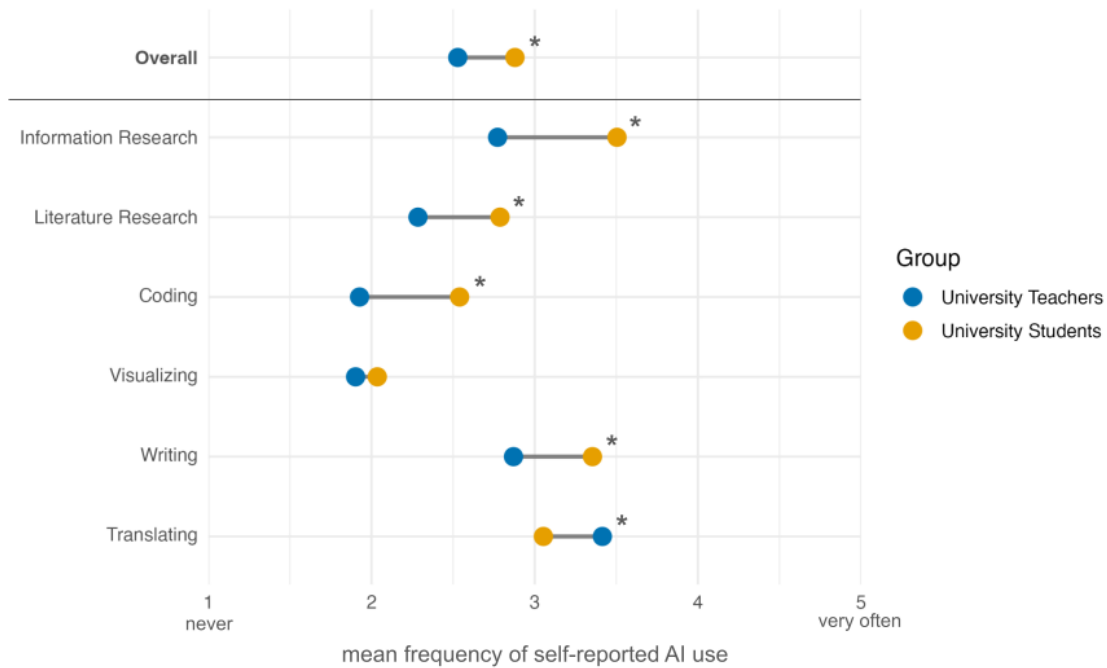

*Note*. Statistical significance of differences at p < .05 is indicated with an asterisk (*).

Overall, students reported using AI more often than teachers, by an average of 0.35 points, indicating a small but significant main effect of group ($b$ = −0.35, $SE$ = 0.09, $p$ < .001, $d$ = 0.36). We found the largest group differences for information research ($b$ = −0.73, $SE$ = 0.15, $p$ < .001, $d$ = 0.75), followed by programming ($b$ = −0.61, $SE$ = 0.15, $p$ < .001, $d$ = 0.63), literature research ($b$ = −0.50, $SE$ = 0.15, $p$ = .001, $d$ = 0.51), and text work ($b$ = −0.48, $SE$ = 0.15, $p$ = .0009, $d$ = 0.50). In contrast, the difference was very small and non-significant for visualization tasks ($b$ = −0.13, $SE$ = 0.15, $p$ = .362, $d$ = 0.14). For translation tasks, the opposite pattern emerged: teachers reported higher AI use than students ($b$ = 0.36, $SE$ = 0.15, $p$ = .013, $d$ = 0.37).

Taken together, students reported using AI more frequently for academic tasks than teachers for five of the six tasks, supporting H1; the only exception is translation tasks, where we find the opposite pattern.

**RQ2: To what degree do students/teachers delegate their academic tasks to AI?**

We conducted a linear mixed-effects regression to examine differences in self-reported degree of AI delegation between university students and teachers, as illustrated in Figure 3. The model included group (students vs. teachers), task type, and their interaction as fixed effects. A random intercept for participants accounted for individual differences in self-reported degree of AI delegation, with $\sigma^2$ = 259.80 (*SD* = 16.12), and a residual variance of $\sigma^2$ = 561.50 (*SD* = 23.70), reflecting within-participant variability across academic tasks.

**Figure 3.**

*Self-reported Degree of AI Delegation of University Teachers and Students.*

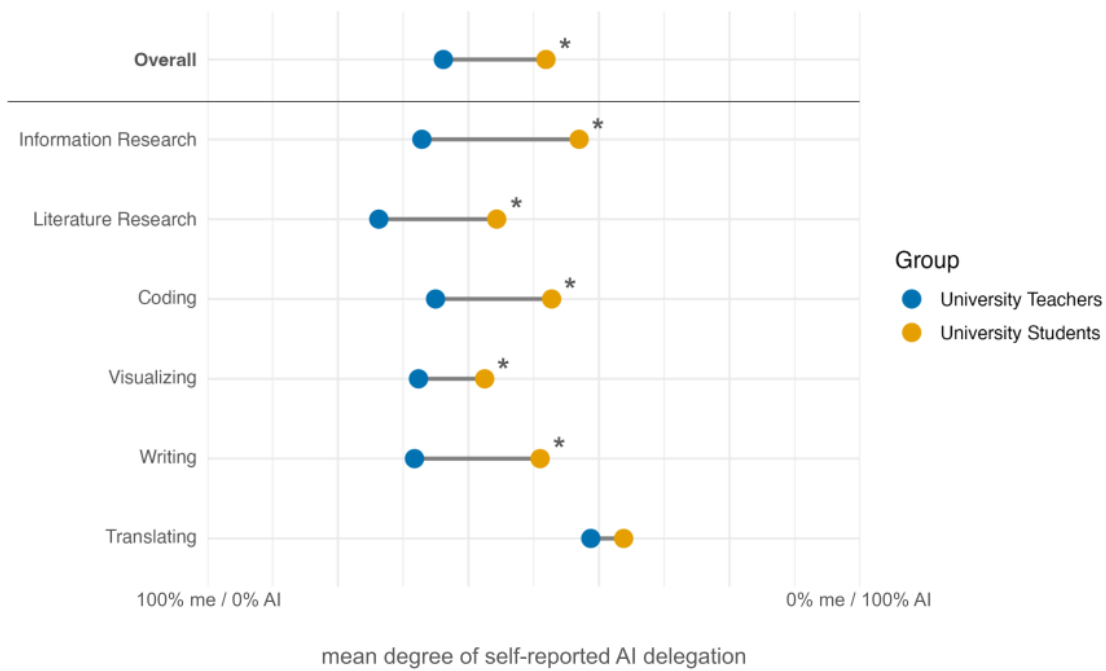

*Note*. Statistical significance of differences at p < .05 is indicated with an asterisk (*).

Overall, students reported delegating tasks to AI to a greater extent than teachers, by an average of 15.72 points, indicating a medium sized and significant main effect of group (*b* = −15.72, *SE* = 2.46, *p* < .001, *d* = 0.66). We found the largest group differences for information research (*b* = −24.09, *SE* = 3.74, *p* < .001, *d* = 1.02), followed by literature research (*b* = −18.06, *SE* = 3.75, *p* < .001, *d* = 0.76), programming (*b* = −17.78, *SE* = 3.80, *p* < .001, *d* = 0.75), and writing (*b* = −19.25, *SE* = 3.74, *p* < .001, *d* = 0.81). In contrast, the difference was smaller for visualizing (*b* = −10.14, *SE* = 3.78, *p* = .007, *d* = 0.43) and translating (*b* = −5.04, *SE* = 3.75, *p* = .180, *d* = 0.21), with the latter not reaching statistical significance.

Taken together, students reported delegating to AI to a greater extent for academic tasks than teachers, supporting H2.

**RQ3.1: How accurately can university students predict teachers' frequency of AI use and vice versa?**

We conducted a linear mixed-effects regression analysis to examine differences in prediction accuracy related to the frequency of AI use, as illustrated in Figure 4. The model included group (teachers vs. students), academic task type, and their interaction as fixed effects. A random intercept for participants accounted for individual differences in prediction accuracy, $\sigma^2$ = 0.06 (*SD* = 0.25), with a residual variance of $\sigma^2$ = 0.34 (*SD* = 0.58), indicating within-participant variability across tasks.

**Figure 4.**

*Self-reports vs. Cross-sample Predictions of Frequency of AI Use among University Teachers and Students.*

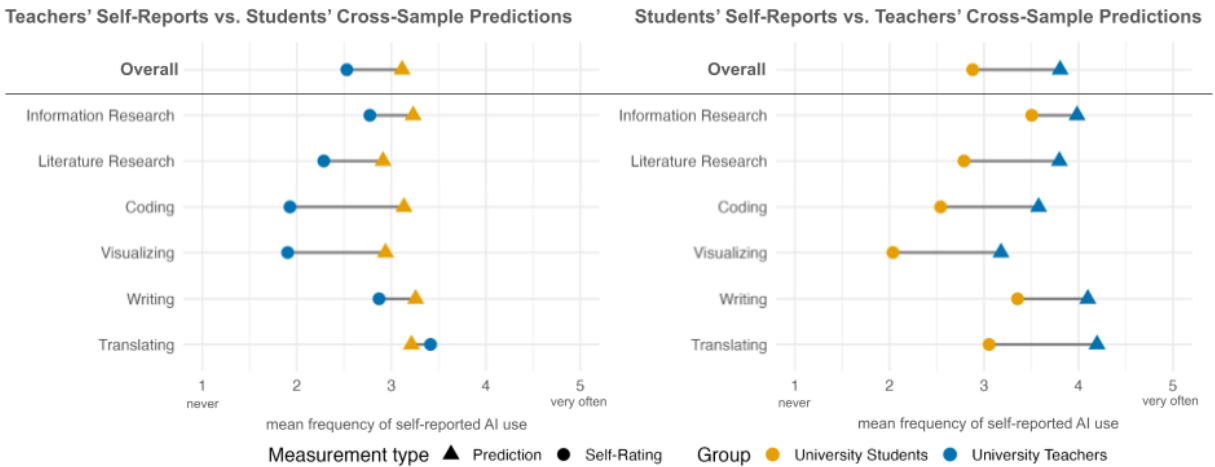

Overall, teachers and students overestimated the other groups' frequency of AI use by 1.02 points, indicating significant prediction inaccuracies of very large effect sizes for both groups, as reflected by the common model intercept across both groups and tasks ($b = 1.02$, $SE = 0.02$, $p < .001$, $d = 1.75$). While both students and teachers were inaccurate in their predictions about the respective other group, we also analyzed whether they differ in the accuracy of their predictions. We observed that students were slightly more accurate in predicting teachers' use than vice versa ($b = -0.093$, $SE = 0.05$, $t(234) = -2.05$, $p = .042$, $d = 0.16$). We note that this effect disappeared when the analysis was conducted with the alternative student sample.

Taken together, we find no evidence that university teachers were more accurate in predicting students' frequency of AI use than vice versa, contradicting H3.1.

**RQ3.2: How accurately can university students predict teachers' degree of AI delegation and vice versa?**

To examine differences in prediction accuracy related to the degree of AI delegation, we conducted a linear mixed-effects regression, as illustrated in Figure 5. The model included group (teachers vs. students), academic task type, and their interaction as fixed effects. A random intercept for participants accounted for individual differences in prediction accuracy, $\sigma^2 = 85.67$ ($SD = 9.26$), with a residual variance of $\sigma^2 = 155.39$ ($SD = 12.47$), indicating within-participant variability across tasks.

**Figure 5.**

*Self-reports vs. Cross-sample Predictions of Degree of AI Delegation among University Teachers and Students.*

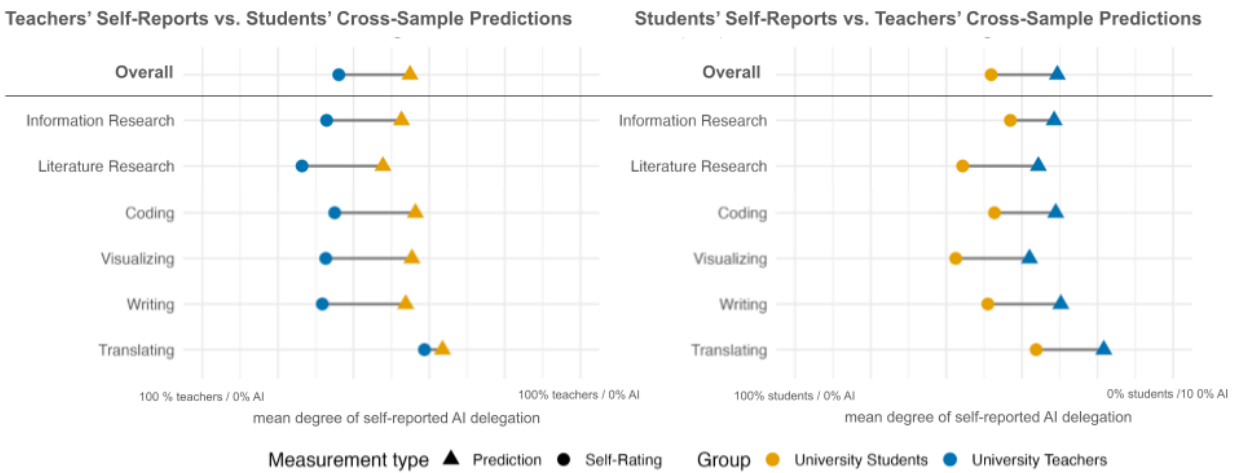

Overall, teachers and students overestimated the other groups' degree of AI delegation by 25.89 points, indicating significant prediction inaccuracies of very large effect sizes for both groups, as reflected by the common model intercept across both groups and tasks ($b = 25.89$, $SE = 0.69$, $p < .001$, $d = 2.08$). While both students and teachers were inaccurate in their predictions about the respective other group, we also analyzed whether they differ in the accuracy of their predictions. We observed that no group was significantly more accurate in predicting the degree of AI delegation of the other group ($b = -2.43$, $SE = 1.38$, $t(233.98) = -1.76$, $p = .079$, $d = 0.20$).

Taken together, we find no evidence that university teachers were more accurate in predicting students' degree of AI delegation than vice versa, contradicting H3.2.

## 5. Discussion

Our results show that, in higher education, students delegate academic tasks to AI more often and to a greater extent than teachers. We also found that both teachers and students overestimate how much the other group uses AI.

As expected, students reported higher and more intensive use of AI tools than teachers across a range of academic tasks, both in terms of frequency and degree of delegation. We observed the largest differences in both measures for information research and literature research, with notable gaps also in programming and writing. In contrast, translation tasks showed the reverse pattern, with teachers reporting greater use and delegation. One possible explanation is that teachers, who are often older and have achieved a high level of professional expertise, rely on well-established problem-solving strategies for academic tasks and are therefore less likely to adopt new technologies. Our findings, in line with prior studies, show that a substantial subset of teachers report very low AI use in both frequency and degree of delegation (Beege et al., 2024; Kiryakova & Angelova, 2023). In contrast, younger people, such as most university students, tend to have greater digital literacy and are quicker to explore emerging tools; however, they may also be less aware of their shortcomings and limitations (Chan & Lee, 2023). Notably, this is one of the first studies to compare AI use between teachers and students using equivalent measurement instruments, allowing for a more direct and reliable assessment of differences in adoption.

Unexpectedly, university teachers and students were equally inaccurate in predicting the other group's AI use across a range of academic tasks, both in terms of frequency and degree of delegation. While generally inaccurate cross-sample predictions can be attributed to the lack of prior experience in higher education in the age of AI, we expected that teachers' predictions would be more accurate than those of students. This expectation was based on the assumption that teachers have richer experience in assessing and observing large and diverse cohorts of students over time. In line with this finding, a qualitative study by Lau and Guo (2023) suggests that teachers struggle to accurately predict students' use of AI. We extend their work by quantitatively comparing teachers' predictions of students' AI use with students' self-reports and identifying a statistically significant difference. Furthermore, the mutual overestimation between groups reveals a striking pattern: teachers consistently overestimate students' AI use, predicting both the frequency and the degree of students' delegation to be very high. This pattern reflects a broader trend toward suspicion of students in higher education, as argued by Macfarlane (2021) and Ross and Macleod (2018). In contrast, students overestimate teachers' AI use by predicting frequencies and degrees of delegation that are similar to their own self-reported values. This pattern reflects a false-consensus effect (Marks & Miller, 1987), where individuals wrongly assume that others behave in ways similar to their own.

## 6. Implications

Our findings indicate a substantial disconnect between teachers' and students' expectations of each other's AI use and the actual application of AI in higher education. This disconnect poses a critical challenge to trusting teacher–student relationships, which are a key element of higher education, linked to learning outcomes, mutual willingness to take risks, and the collaborative exploration of new technologies (Cavanagh et al., 2018; Felten et al., 2023; Luo, 2025).

To foster mutual trust in the age of AI, teachers and students should therefore engage in open discussions about the role and application of AI in academic contexts, particularly to address misconceptions about its use. Following Luo's (2025) argument, a two-way transparency is required. Teachers, like students, should publicly acknowledge the AI assistance they use in their work, and in some cases, even provide supporting documentation, such as chat records, to ensure accountability at the same level expected from students.

## 7. Limitations and future research

While this study provides new insights into the use of AI in higher education, several limitations should be acknowledged when interpreting the findings.

First, all data in this study stem from self-reports, reflecting participants' subjective accounts. Such measures may be affected by biases, including inaccurate recall, social desirability, or misinterpretation of questions (Corneille & Gawronski, 2024). This issue is particularly relevant in domains where social desirability bias can occur, as respondents may adjust their answers to align with perceived norms (Fisher & Katz, 2000). In the context of AI use in academia, ongoing debates about its legitimacy may influence how openly participants report their behavior. We attempted to mitigate the impact of social desirability by informing all participants that their responses would be treated confidentially and anonymized, and by emphasizing that there were no "right" or "wrong" answers. These measures could have reduced, but may not have entirely eliminated, the tendency to provide socially acceptable responses. Future research could complement self-report data with objective behavioral measures, such as actual usage logs from AI tools, system activity records, or passive data collection methods that capture AI interaction patterns in real-time. Combining subjective and objective data sources would enable a more accurate understanding of AI use in academic contexts, helping to determine whether discrepancies exist between reported and actual behavior.

Second, to ensure that our surveys were relevant to teachers and students from all academic fields, we selected domain-unspecific academic tasks and phrased them to be tangible for participants across diverse disciplines. While this approach may only partly reflect participants' everyday tasks and does not capture the varying importance of specific tasks in different disciplines, we see it as a necessary first step for gaining insights into cross-sample predictions about AI use in academic contexts. Future research should build on this foundation by incorporating discipline-specific tasks and exploring how beliefs and usage patterns may vary across different academic domains.

Third, and relatedly, the study did not aim to test differences across fields of study. Therefore, we did not recruit sufficiently large samples within each field. A promising direction for future research is to examine more robustly whether fields of study differ in their actual AI use and beliefs about AI use. This is particularly relevant as prior work suggests that disciplines with a higher prevalence of AI tools or established digital practices, such as engineering and the natural sciences, may report greater use and delegation compared to fields where such tools are less common (Von Garrel & Mayer, 2023).

Fourth, we deviated from our pre-registration by not recruiting a full convenience sample of university students ($n$ = 47) under the same circumstances as the teacher sample. To address this, we conducted a second wave of data collection via Prolific. This method carries the risk of recruiting a student population that is less representative and potentially more digitally affluent. However, the Prolific student sample reported values comparable to prior work (Reiter et al., 2025; Von Garrel & Mayer, 2023). A detailed comparison of the incomplete convenience sample and the Prolific sample is provided in the Supplementary Information. Recruiting via Prolific also introduced differences in remuneration between groups. Teachers could earn donations made in their name, whereas students received direct monetary payments. This difference in compensation may have influenced motivation and mediated some responses, although there is no clear evidence that it meaningfully affected the results. We consider these differences minor since the remuneration was similar in size. Future research should address this limitation by using studies with identical incentives for all participant groups.

Lastly, we acknowledge that this study represents a single point in time. As teachers and students gain more experience with AI, their cross-sample predictions about the other group's AI use may become more accurate. Future research could therefore build on our results by including measures of perceived trustworthiness and examining its relationship with prediction accuracy over longer periods. Directly testing this correlation across repeated measurements would yield valuable insights into how experience and trust interact in shaping accurate cross-group judgments.

## 8. Conclusion

In sum, this study makes three key contributions to research on AI in higher education. First, we extend prior work by systematically examining university teachers as a distinct sample, using equivalent measures that allow direct comparison with students. Second, we go beyond usage rates by capturing both the frequency and the degree of delegation of academic tasks to AI, thereby providing a more nuanced understanding of the adoption process. Third, we offer some of the first empirical insights into mutual predictions of AI use between teachers and students, employing incentivized cross-sample predictions to link these beliefs to issues of trust. Together, these contributions address important gaps in the existing literature and highlight how misjudgments about AI use may influence adoption patterns and the quality of teacher-student relationships in the era of AI.

Advancing this line of inquiry will not only deepen our understanding of the dynamics between trust, perception, and technology use in higher education but also inform policies and practices that promote constructive and equitable integration of AI tools across diverse learning environments.

**Refrences**